\documentclass[aps,prl,twocolumn,superscriptaddress,amsmath,amssymb]{revtex4}
\usepackage{amssymb}
\usepackage{amsmath}
\usepackage{amsfonts}
\usepackage{epsfig}
\usepackage{color}
\usepackage{xcolor}
\usepackage{graphics, graphicx}
\usepackage{bbold}
\usepackage{psfrag}
\usepackage{mathcomp}
\usepackage{subfigure}
\usepackage{verbatim}
\usepackage{color}
\usepackage[colorlinks,citecolor=blue]{hyperref}
\def\bm#1{\mathbf{#1}}

\begin{document}

\title{Tuning Feshbach resonances in cold atomic gases with inter-channel coupling}

\author{Tian-Shu Deng}
\affiliation{Key Laboratory of Quantum Information, University of Science and Technology of China, Chinese Academy of Sciences, Hefei, Anhui, 230026, China}
\affiliation{Synergetic Innovation Center of Quantum Information and Quantum Physics, University of Science and Technology of China, Hefei, Anhui 230026, China}
\author{Wei Zhang}
\email{wzhangl@ruc.edu.cn}
\affiliation{Department of Physics, Renmin University of China, Beijing 100872, China}
\affiliation{Beijing Key Laboratory of Opto-electronic Functional Materials and Micro-nano Devices,
Renmin University of China, Beijing 100872, China}
\author{Wei Yi}
\email{wyiz@ustc.edu.cn}
\affiliation{Key Laboratory of Quantum Information, University of Science and Technology of China, Chinese Academy of Sciences, Hefei, Anhui, 230026, China}
\affiliation{Synergetic Innovation Center of Quantum Information and Quantum Physics, University of Science and Technology of China, Hefei, Anhui 230026, China}

\date{\today}
\begin{abstract}
We show that the essential properties of a Feshbach resonance in cold atomic gases can be tuned by dressing the atomic states in different scattering channels through inter-channel couplings. Such a scheme can be readily implemented in the orbital Feshbach resonance of alkaline-earth-like atoms by coupling hyperfine states in the clock-state manifolds. Using $^{173}$Yb atoms as an example, we find that both the resonance position and the two-body bound-state energy depend sensitively on the inter-channel coupling strength, which offers control parameters in tuning the inter-atomic interactions. We also demonstrate the dramatic impact of the dressed Feshbach resonance on many-body processes such as the polaron to molecule transition and the BCS-BEC crossover.
\end{abstract}
\maketitle

\begin{figure}[tbp]
\center{\includegraphics[width=1\linewidth]{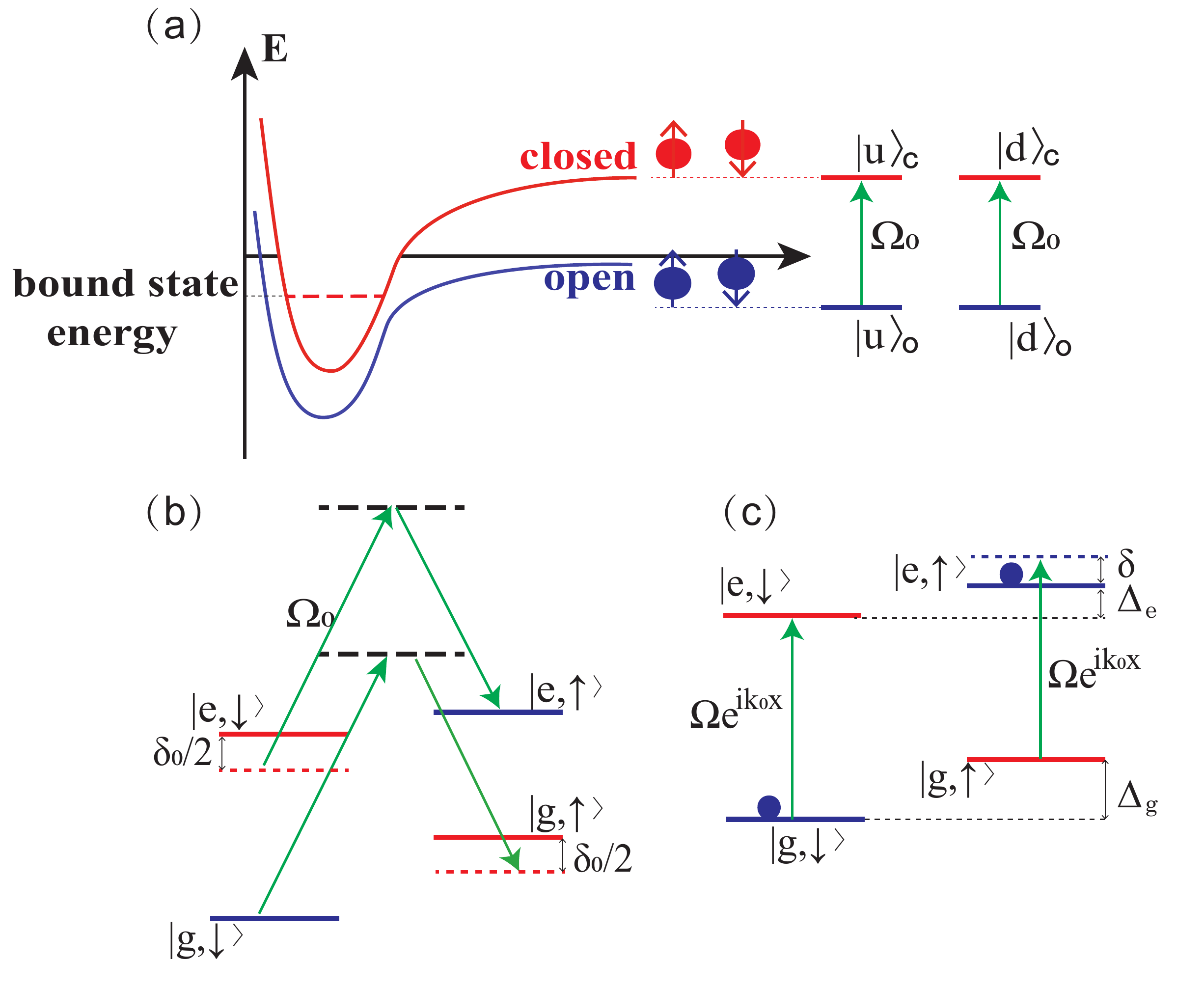}}
\caption{ (a) Left: illustration of the scattering channels in a typical $s$-wave FR. Right: dressing the FR with inter-channel couplings. (b) Raman scheme for the inter-channel coupling in an OFR. (c) Rabi scheme for the inter-channel coupling in an OFR. The labels for the atomic states, the coupling parameters, and the detunings are defined in the text.}
\label{fig:fig1}
\end{figure}

\emph{Introduction}.--
Feshbach resonance (FR) has been a key element in the toolbox of quantum control in cold atomic gases~\cite{FRreview}. By making the strongly-interacting regime accessible, FR enables the preparation and investigation of strongly-correlated many-body quantum states in the highly controllable environment of cold atoms. The essential properties of FRs in cold atomic gases can be grasped by considering a two-channel scattering process, in which a scattering resonance occurs as a bound molecular state in the so-called closed channel crosses the continuum threshold of the open channel (see Fig.~\ref{fig:fig1}(a)). As the interaction potentials associated with both scattering channels typically depend on the internal states of atoms, external magnetic or optical fields can be applied to shift the potentials and tune the inter-atomic scattering length.

Previous studies have shown that FRs can be modified either by dressing the molecular bound state in the closed channel~\cite{dressFR1,dressFR2,dressFR3,dressFR4,dressFR5,dressFR6,dressFR7,dressFR8}, or by coupling different atomic states in the open channel~\cite{dressFRopen1,dressFRopen2,dressFRopen3,dressFRopen4,dressFRopen5}. Under these situations, the resonance position as well as the atomic scattering length can be tuned by additional parameters. In principle, inter-channel couplings between atomic states should also modify the resonant scattering by shifting the relative position between the continuum thresholds of the scattering channels. However, in the conventional magnetic FR of alkali-metal atoms, the open- and the closed-channel thresholds are far-detuned, such that the scattering states in the closed channel are not accessed in the low-energy scattering. This is not the case in the recently discovered orbital Feshbach resonance (OFR) in alkaline-earth-like atoms~\cite{ren1,ofrexp1,ofrexp2}, where the continuum thresholds of the two scattering channels are close to one another.  This opens up the interesting possibility of dressing FRs by inter-channel couplings.

A typical OFR in alkaline-earth-like atoms involves four hyperfine states in the ground $^1S_0$ (referred to as the $|g\rangle$ orbital) and the metastable $^3P_0$ (the $|e\rangle$ oribtal) manifolds~\cite{ren1}. In an OFR, the open channel corresponds to one atom in $|g\downarrow\rangle$ and the other in $|e\uparrow\rangle$, the closed channel corresponds to one in $|g\uparrow\rangle$ and the other in $|e\downarrow\rangle$. Here, $|\downarrow\rangle$ and $|\uparrow\rangle$ represent two different nuclear-spin states in the hyperfine manifolds of the clock states $^1S_0$ and $^3P_0$. As illustrated in Fig.~\ref{fig:fig1}(b)(c), the inter-channel coupling can be implemented either by imposing Raman lasers coupling the nuclear-spin states in the same manifold (the Raman scheme), or by directly driving the clock transition $^1S_0$-$^3P_0$ (the Rabi scheme). While coupling the clock states typically require ultra-stable and high-power lasers, which can give rise to additional heating due to the photon recoil, these difficulties should be manageable with existing techniques, particularly in light of the recent experimental realization of synthetic spin-orbit coupling in alkaline-earth-like atoms~\cite{ye2016old,ye2016,fallani2016,gyuboong}. Note that while the inter-channel couplings also realizes synthetic SOC within the clock-state manifolds, as has recently been experimentally realized, the dressing of the FR and the modification of the resonance properties are related to the shifting of the continuum thresholds of the scattering channels, rather than the momentum transfer of the SOC. With the inter-channel couplings, we will show that the resonance position as well as the scattering length of the OFR are drastically modified by the coupling strength. For example, in $^{173}$Yb atoms, given a typical coupling strength, the shift in the resonance position can be on the order of the resonance width~\cite{ofrwidth}, which gives rise to resonant interactions even at zero magnetic field.
Thus, by providing additional control parameters over the few-body and the associated many-body states across the dressed FR, our scheme not only holds the potential of extending the flexibility of FRs in cold atomic gases, but also has immediate implications for the quantum simulation using alkaline-earth-like atoms near the orbital FR. These include new routes toward enhancing Kondo coupling~\cite{kondo1,kondo2,kondo3,kondo4,kondo5,kondo6}, as well as the interesting possibility of investigating many-body localization~\cite{mbl,mbl2} or the Floquet dynamics~\cite{floquet} by introducing spatial or temporal modulation of interaction potentials~\cite{ChinFRmod}. In the following, before discussing the impact of inter-channel couplings on the few- and many-body properties of OFR, we first give a general description of the dressed FR with a minimal two-channel model.

\emph{Model}.--
We consider a two-channel model for the scattering of two atoms with mass $m$, where the two atomic internal states in the closed channel labeled by $\{| u \rangle_c,| d \rangle_c \}$ and the ones in the open channel $\{| u \rangle_o,| d \rangle_o \}$ are dressed by inter-channel couplings. The non-interacting Hamiltonian of the relative motion is
\begin{align}\label{f1H0}
    H_{0} & =\left(-\frac{\hbar^{2}\nabla^{2}}{m}+\delta_{0}\right)|cc\rangle\langle cc|
    - \frac{\hbar^{2}\nabla^{2}}{m} |oo\rangle\langle oo|\nonumber\\
     & +\left(-\frac{\hbar^{2}\nabla^{2}}{m}+\frac{\delta_{0}}{2}\right)
     \left(|co\rangle\langle co|+|oc\rangle\langle oc| \right)\nonumber\\
     &+\sum_{n=\{u,d\}}\Omega_{0} \left(|n\rangle_{c}\langle n|_o+H.c.\right)
\end{align}
where $|cc\rangle=|d \rangle_c|u\rangle_c$, $|oo\rangle=|d\rangle_o |u\rangle_o$, $|co\rangle=|d\rangle_c|u\rangle_o$, and $|oc\rangle=|d\rangle_o|u\rangle_c$ form the Hilbert space of two-body wave functions. $\delta_0/2$ gives the single-particle energy detuning between the scattering channels, and $\Omega_0$ is the inter-channel coupling strength. Without loss of generality, the interaction Hamiltonian can be written in the form of the Huang-Yang pseudo-potential
\begin{align}\label{f1Hint}
H_{\rm int}=\frac{4\pi\hbar^{2}}{m}\sum_{i,j}a_{ij}|ii\rangle\langle jj| \delta(\bm{r})\frac{\partial}{\partial r}(r\cdot),
\end{align}
where $\bm{r}$ is the relative coordinate, and $a_{ij}$ ($i,j=\{o,c\}$) is the corresponding $s$-wave scattering length.

Under the inter-channel coupling, the single-particle incident scattering states become $|n\rangle_1=\cos\theta|n\rangle_c-\sin\theta|n\rangle_o$ and $|n\rangle_2=\sin\theta|n\rangle_c+\cos\theta|n\rangle_o$, where the indices $(1,2)$ label the new incident scattering channels, and $\tan\theta=\left(\delta_0/4+\sqrt{(\delta_0/4)^2+\Omega_0^{2}}\right)/\Omega_0$. The non-interacting Hamiltonian $H_0$ is diagonal under the basis $\{|d\rangle_\alpha|u\rangle_\beta\}$ ($\alpha,\beta=1,2$), with $H_{0}=\sum_{\alpha,\beta}(-\hbar^{2}\nabla^{2}/{m}+\epsilon_{\alpha}+\epsilon_{\beta})|\alpha\beta\rangle\langle\alpha\beta|$, where $|\alpha\beta\rangle=|d\rangle_\alpha|u\rangle_\beta$, and $\epsilon_{1,2}=\delta_{0}/4\mp\sqrt{(\delta_{0}/4)^{2}+\Omega_{0}^{2}}$. The scattering wave function can then be written as
\begin{align}\label{sw}
    |\Psi(\bm{r})\rangle &=\left[e^{i\bm{k}\cdot\bm{r}}+f_{11}(\bm{k})\frac{e^{ikr}}{r}\right]|d\rangle_1 |u\rangle_1\nonumber\\
    &+\sum_{\alpha,\beta\neq (1,1)}f_{\alpha\beta}(\bm{k})\frac{e^{-\kappa_{\alpha\beta}r}}{r}|d\rangle_\alpha|u\rangle_\beta,
\end{align}
where $\kappa_{\alpha\beta}=\sqrt{m\Delta_{\alpha\beta}/\hbar^{2}-k^{2}}$, $\Delta_{\alpha\beta}=\epsilon_{\alpha}+\epsilon_{\beta}-2\epsilon_1$, and $f_{\alpha\beta}$ is the scattering amplitude of the corresponding channel. $\hbar\bm{k}$ is the relative momentum with respect to the scattering threshold $2\epsilon_1$.

Substituting Eq.~(\ref{sw}) into the Schr\"odinger's equation $(H_{0}+H_{\rm int}-{\hbar^{2}k^{2}}/{m}-2\epsilon_1)|\Psi(\bm{r})\rangle=0$, we get a set of coupled equations for the scattering amplitudes. We may then extract the low-energy scattering length from $f_{11}(\bm k)$, which belongs to the lowest-energy scattering channel
\begin{align}\label{f11}
    a_{s}^{(11)}&=-\underset{\bm{k}\rightarrow0}{\lim}f_{11}(\bm{k})\nonumber\\
    &=-\frac{a_{1}\sqrt{R}(2+4\sqrt{2}\cot^{2}2\theta)-a_{4}}{2\sqrt{2}R a_{1}+(-\sqrt{2R}a_{3}-2\sqrt{R}a_{2})+1}
\end{align}
where $a_{1}=(a_{cc}a_{oo}-a_{co}^{2})\sin^{2}\theta\cos^{2}\theta$, $a_{2}=(a_{cc}+a_{oo}-2a_{co})\sin^{2}\theta\cos^{2}\theta$, $a_{3}=a_{cc}\sin^{4}\theta+a_{oo}\cos^{4}\theta+2a_{co}\sin^{2}\theta\cos^{2}\theta$, $a_{4}=a_{cc}\cos^{4}\theta+a_{oo}\sin^{4}\theta+2a_{co}\sin^{2}\theta\cos^{2}\theta$ and $R=m\sqrt{\delta_0^{2}/4+4\Omega_0^{2}}/\hbar^2$. Note we have assumed $a_{co}=a_{oc}$ in the derivation. The scattering resonance occurs when $a_{s}^{(11)}$ diverges. As the denominator of Eq.~(\ref{f11}) is dependent on $\Omega_0$ and $\delta_0$, both the scattering length and the resonance location should depend on these parameters.

\emph{Implementation}.--
With long-lived excited states and flexible controls over the clock states, OFR in alkaline-earth-like atoms offers a natural platform for the realization of the dressed FR. As illustrated in Fig.~\ref{fig:fig1}, the inter-channel couplings can be achieved by either the Raman scheme (Fig.~\ref{fig:fig1}(b)) or the Rabi scheme (Fig.~\ref{fig:fig1}(c)). In particular, when the Raman lasers in the Raman scheme are co-propagating, the relative motion of the non-interacting system corresponding to the setup in Fig.~\ref{fig:fig1}(b) can be described by the minimum model Eq.~(\ref{f1H0}), with $(u,d)$ corresponding to the so-called orbital degrees of freedom $(g,e)$. $\Omega_0$ is given by the effective Rabi frequency of the Raman process, and the detuning $\delta_0$ is given by the differential Zeeman shift of the clock-state manifolds~\cite{zeemanshift1,zeemanshift2}. In an OFR, the two-body interactions at the short range occur either in the electronic spin-singlet and nuclear spin-triplet channel, with $s$-wave scattering length $a_-$; or in the electronic spin-triplet and nuclear spin-singlet channel, with scattering length $a_+$~\cite{ofr1,ofr2,ofr3}. Thus, the scattering lengths $a_{\pm}$ associated with these short-range potentials are related to the scattering lengths in Eq.~(\ref{f1Hint}) as $a_{cc}=a_{oo}=(a_{+}+a_{-})/2$, $a_{co}=(a_{+}-a_{-})/2$. In this case, Eq.~(\ref{f11}) can be directly applied to describe the dressed OFR.

In the more general case of finite photon recoils in the Raman process, or in the case of the Rabi scheme, where photon recoils are inevitable, the minimal model discussed above becomes inadequate. Furthermore, due to the narrow line-width of the states in the $^3P_0$ manifold, the heating in the laser-coupling process should be significantly reduced in the Rabi scheme, which makes it more appealing compared to the Raman scheme. In the following, we will focus on the dressed OFR under the Rabi scheme, using $^{173}$Yb atoms as a concrete example.

\emph{Dressed resonance in the Rabi scheme}.--
Due to the inevitable momentum transfer in the Rabi scheme, the relative and the center-of-mass motion of the scattering states are coupled, which makes the characterization of the scattering process rather cumbersome. However, one can still identify the resonance from the two-body bound-state threshold. We start from the non-interacting Hamiltonian in the second quantized form
\begin{align}\label{H0}
H'_0&=\sum_{\bm{k},j}\left(\epsilon_{\bm{k}}+\eta_{j}\frac{\hbar^2 k_0 k_x}{m}\right)a^{\dag}_{j,\downarrow,\bm{k}}a_{j,\downarrow,\bm{k}}\nonumber\\
&+ \sum_{\bm{k},j}\left(\epsilon_{\bm{k}}+\eta_{j}\frac{\hbar^2 k_0 k_{x}}{m}+\Delta_j\right)a^{\dag}_{j,\uparrow,\bm{k}}a_{j,\uparrow,\bm{k}}\nonumber\\
&+\Omega\sum_{\bm{k},\sigma}(a^{\dag}_{g,\sigma,\bm{k}}a_{e,\sigma,\bm{k}}+H.c.)
\end{align}
where $a^{\dag}_{j,\sigma,\bm{k}}$ ($a_{j,\sigma,\bm{k}}$)  creates (annihilates) an atom in the corresponding pseudo-spin state $e^{-i\eta_j \frac{k_0 x}{2}}|j,\sigma\rangle$ ($j=\{g,e\}$, $\sigma=\{\uparrow,\downarrow\}$) with momentum $\bm{k}$. Here, $\eta_{g/e}=\pm$, $k_0$ and $\Omega$ are respectively the wave vector and the Rabi frequency of the coupling laser, $\epsilon_{\bm{k}}=\hbar^2 k^2/2m$, and the Zeeman shift of the state $|j\rangle$ ($j=\{g,e\}$ given by $\Delta_{j}=g_j\mu_B B$, with $\mu_B$ the Bohr magneton, $g_j$ the Lande factor, and $B$ the external magnetic field. We define the helicity operators $a_{\pm,\sigma,\bm{k}}=\cos\theta^{\pm}_{\sigma,\bm{k}}a_{g,\sigma,\bm{k}}+\sin\theta^{\pm}_{\sigma,\bm{k}}a_{e,\sigma,\bm{k}}$,
where
$\sin\theta_{\sigma,\bm{k}}^{+}=\Omega/\sqrt{\Omega^{2}+(\hbar^2 k_0 k_{x}/2m-\xi_{\sigma})^{2}}$, $\theta^{-}_{\sigma,\bm{k}}=\theta^{+}_{\sigma,\bm{k}}+\pi/2$,
with $\xi_{\bm{k},\uparrow}=\sqrt{\Omega^{2}+(\hbar^2 k_0 k_{x}/2m-\delta/2)^{2}}+\delta/2$, $\xi_{\bm{k},\downarrow}=\sqrt{\Omega^{2}+(\hbar^2 k_0 k_{x}/2m)^{2}}$, and $\delta=\Delta_g-\Delta_e$. The single-particle Hamiltonian can then be written as $H'_{0}=\sum_{\bm{k},\nu,\sigma}E_{\bm{k},\sigma}^{\nu}a_{\nu,\sigma,\bm{k}}^{\dagger}a_{\nu,\sigma,\bm{k}}$, with $E_{\bm{k},\downarrow}^{\pm}  =\epsilon_{\bm{k}}\pm\sqrt{\Omega^{2}+(\hbar^2 k_0 k_{x}/2m)^{2}}$, and $E_{\bm{k},\uparrow}^{\pm} =\epsilon_{\bm{k}}+\Delta_{e}+\delta/2\pm\sqrt{\Omega^{2}+(\hbar^2 k_0 k_{x}/2m-\delta/2)^{2}}$. The interaction Hamiltonian is~\cite{ren1}
\begin{equation}\label{Hint}
H'_{\rm int}=\frac{g_{+}}{2}\sum_{\bm{q}}A_{+}^{\dagger}(\bm{q})A_{+}(\bm{q})+\frac{g_{-}}{2}\sum_{\bm{q}}A_{-}^{\dagger}(\bm{q})A_{-}(\bm{q}),
\end{equation}
where we have $A_{+}(\bm{q})=\sum_{\bm{k}}(a_{g,\uparrow,\bm{q-k}}a_{e,\downarrow,\bm{k}}-a_{g,\downarrow,\bm{q-k}}a_{e,\uparrow,\bm{k}})$ and $A_{-}(\bm{q})=\sum_{\bm{k}}(a_{g,\uparrow,\bm{q-k}}a_{e,\downarrow,\bm{k}}+a_{g,\downarrow,\bm{q-k}}a_{e,\uparrow,\bm{k}})$. The interaction strength $g_\pm$ are related to the scattering lengths $a_{\pm}$ as: $1/g_{\pm}=1/g^p_{\pm}-\sum_{\bm{k}}1/2\epsilon_{\bm{k}}$ and $g^p_{\pm}=4\pi\hbar^{2}a_{\pm}/m$.

\begin{figure}[tbp]
\center{\includegraphics[width=1\linewidth]{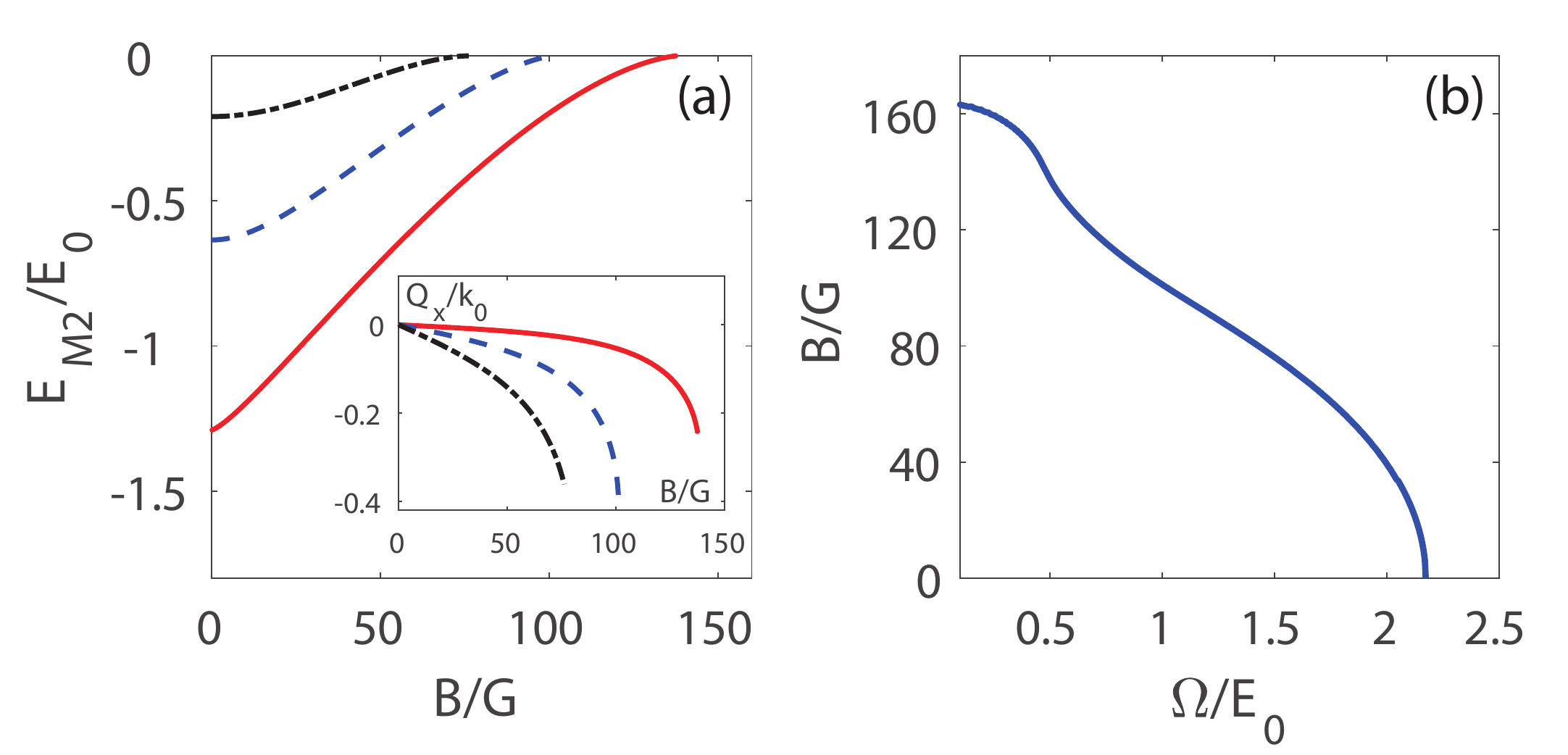}}
\caption{(a) The lowest two-body bound-state energies $E_{M2}$ as functions of the magnetic field $B$ for the dressing parameters $\Omega/E_0=0.5$ (red solid), $\Omega/E_0=1$ (blue dashed), and $\Omega/E_0=1.5$ (dash-dotted), respectively. The corresponding center-of-mass momenta are aligned along the $x$ direction with magnitude $Q_x$ shown in the inset. (b) The two-body resonance point where the bound-state energy reaches the threshold in the $\Omega$--$B$ plane. Here $k_0$ is the wave vector of the $578$nm clock transition $^{1}S_{0}\rightarrow{}^{3}P_{0}$ in $^{173}\rm{Yb}$, and we define the unit of energy through $E_0=\hbar^2k_0^2/2m$. For concreteness, we have taken the parameters of $^{173}\rm{Yb}$ for our calculations, with $a_s^-=219.5a_0$, $a_s^+=1900a_0$, $g_g\mu_B=2\pi\hbar\times 207.15$Hz/G, $g_e\mu_B=2\pi\hbar\times93.78$Hz/G, where $a_0$ is the Bohr radius~\cite{ofrexp1,ofrexp2,zeemanshift1,zeemanshift2}.}
\label{fig:fig2}
\end{figure}

The wave function of the two-body bound state can be written as
\begin{equation}\label{M2}
    |M_{2}\rangle_{\bm{Q}} =\sum_{\bm{k}}\sum_{\mu,\nu=\pm}\psi_{\bm{k}}^{\mu\nu}a_{\mu,\uparrow,\bm{Q}-\bm{k}}^{\dagger}a_{\nu,\downarrow,\bm{k}}^{\dagger}|\rm{vac}\rangle.
\end{equation}
with the bound-state wave function $\psi_{\bm k}^{\mu\nu}$. From the Schr\"{o}dinger's equation $(H'_{0}+H'_{\rm int})|M_{2}\rangle_{\bm{Q}}=(E_{M2}+E^{(2)}_{\rm th})|M_{2}\rangle$, we derive the closed equation for the two-body bound state as $\det{G}=0$ with
\begin{equation}\label{CEM2}
    G=\left[
    \begin{array}{cc}
\frac{1}{2}(\frac{1}{g{}_{+}}+\frac{1}{g{}_{-}})-F_{11}& \frac{1}{2}(\frac{1}{g{}_{+}}-\frac{1}{g{}_{-}})-F_{12}\\
\frac{1}{2}(\frac{1}{g{}_{+}}-\frac{1}{g{}_{-}})-F_{21}& \frac{1}{2}(\frac{1}{g{}_{+}}+\frac{1}{g{}_{-}})-F_{22}
\end{array}\right],
\end{equation}
and
\begin{equation}\label{Fmn}
F_{mn}=\sum_{\bm{k}}\sum_{\mu,\nu}\frac{f_{\mu\nu}^{m}(\bm{Q},\bm{k})f_{\mu\nu}^{n}(\bm{Q},\bm{k})}{(E_{M2}-E_{\bm{Q-k},\uparrow}^{\mu}-E_{\bm{k},\downarrow}^{\nu})},
\end{equation}
and $f_{\mu\nu}^{1}=\sin\theta_{\downarrow,\bm{k}}^{\nu}\cos\theta_{\uparrow,\bm{Q-k}}^{\mu}$, $f_{\mu\nu}^{2}=\cos\theta_{\downarrow,\bm{k}}^{\nu}\sin\theta_{\uparrow,\bm{Q-k}}^{\mu}$.
The ground state can be solved by minimizing $E_{M2}$ with respect to the center-of-mass momentum ${\bf Q}$. Here the two-body threshold energy $E_{\rm th}^{(2)}=\epsilon^0_{\uparrow}+\epsilon^0_{\downarrow}$, with $\epsilon^0_{\sigma}=\min(E_{\bm{k},\sigma}^-)$. Note that $\epsilon^0_{\uparrow}=\epsilon^0_{\downarrow}$ at $B=0$.

As shown in Fig.~\ref{fig:fig2}, both the bound-state energy and the bound-state threshold ($E_{M2}=0$) are functions of the magnetic field $B$ and the dressing parameter $\Omega$. In particular, as demonstrated in Fig.~\ref{fig:fig2}(b), the bound-state threshold can be reached by tuning $\Omega$ even at zero magnetic field, which suggests a scattering resonance by tuning the dressing parameter alone.

\begin{figure}[tbp]
\center{\includegraphics[width=1\linewidth]{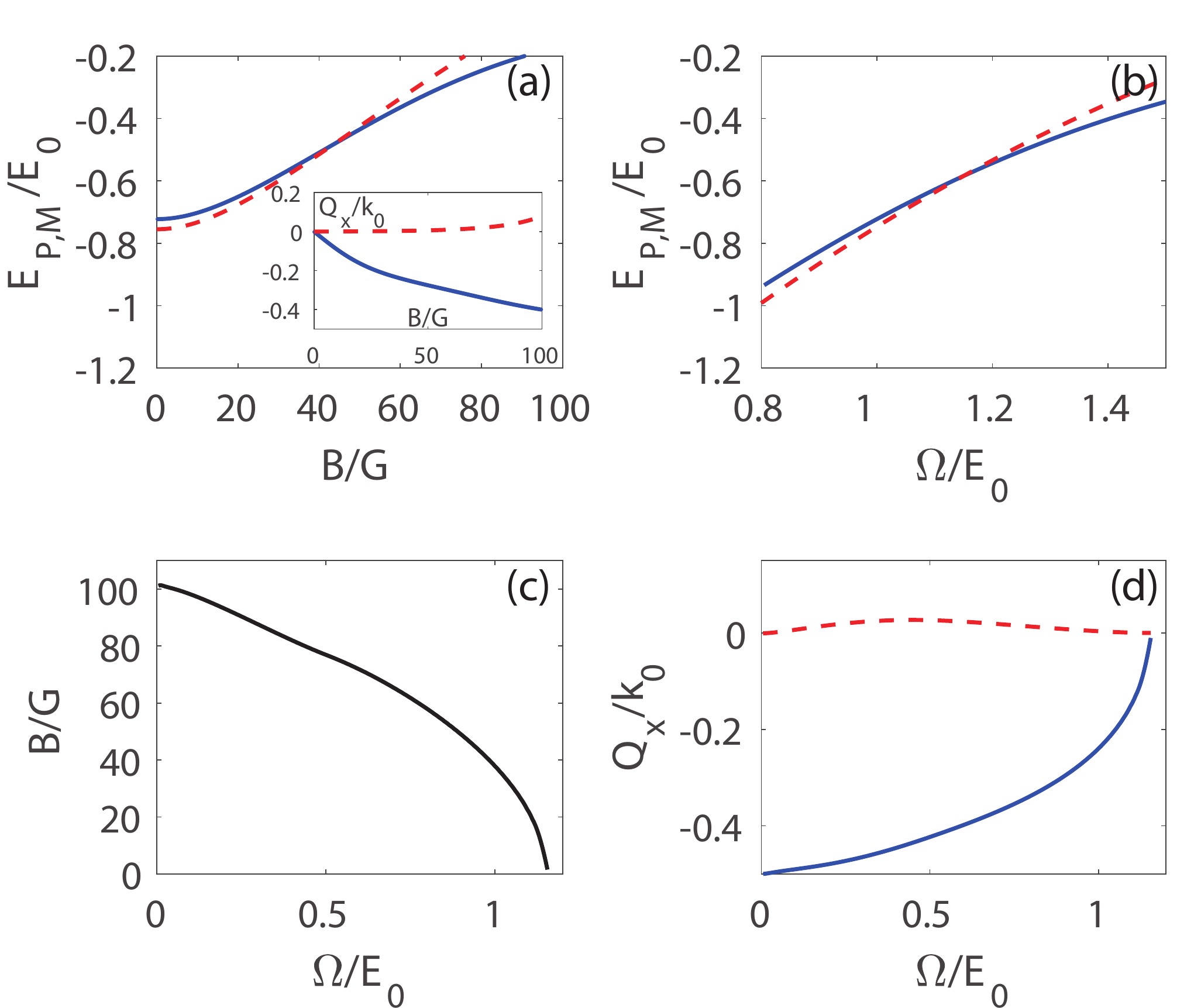}}
\caption{(a) Molecule (red dashed) and polaron (blue solid) energies as functions of the magnetic field $B$ at $\Omega/E_0=1$. The inset shows the center-of-mass momentum $Q_x$ of the polaron (blue solid line) and the molecule (red dashed line), respectively. (b) Molecule (red dashed line) and polaron energies (blue solid line) as functions of $\Omega$ with $B=0$. Here the center-of-mass momenta of both states are zero. (c) Polaron-molecule transition on the $\Omega$-$B$ plane. (d) The center-of-mass momentum $Q_x$ at the transition point for polaron (blue solid line) and molecule (red dashed line), respectively. Here the Fermi energy relative to the single-particle-dispersion minimum $\epsilon^0_{\downarrow}$  is taken as $(E_{F}-\epsilon^0_{\downarrow})/E_0=0.25$. The atomic parameters of $^{173}$Yb are shown in the caption of Fig.~\ref{fig:fig2}.}
\label{fig:fig3}
\end{figure}

\emph{Impurity problem}.--
The dressing of the OFR on the few-body level can lead to various interesting many-body effects. As an exemplary case intervening few- and many-body scenarios, we first study the impurity problem where an impurity atom in the $|\uparrow\rangle$ state interacts attractively with a Fermi sea of $N$ atoms in the $|\downarrow\rangle$ state. In the presence of the coupling laser in the Rabi scheme, the single-particle eigen states for both the majority atoms and the impurity are the helical states. The many-body ground state of such a system can undergo a polaron to molecule transition as the interaction strength increases. In the absence of the coupling laser, it has been shown that the transition occurs at a given magnetic field~\cite{AEpolaron}. With a coupling laser dressing the OFR, we will show that this is no longer the case, as the transition becomes dependent on the dressing parameter $\Omega$.

The molecule ($|M\rangle_{\bm Q}$) and the polaron ($|P\rangle_{\bm Q}$) states can be described using the Chevy-type ansatz~\cite{polaronreview1,polaronreview2}
\begin{align}\label{eqn:MandP}
&|M\rangle_{\bm{Q}}=\sum_{\mu,\nu}\sum_{E_{\bm{k},\downarrow}^{\nu}>E_{F}}\phi_{\bm{k}}^{\mu\nu}a_{\mu,\uparrow,\bm{Q-k}}^{\dagger}a_{\nu,\downarrow,\bm{k}}^{\dagger}|\rm{FS}\rangle_{N-1},\\
   &|P\rangle_{\bm{Q}}=\sum_{\mu}\psi_{\bm{Q}}^{\mu}a_{\mu,\uparrow,\bm{Q}}^{\dagger}|\rm{FS}\rangle_{N}\nonumber\\
 &+\sum_{\mu\nu\lambda}\sum_{\substack{E_{\bm{q},\downarrow}^{\mu}<E_{F}\\
E_{\bm{k},\downarrow}^{\nu}>E_{F}}
}\psi_{\bm{k,q}}^{\mu\nu\lambda}a_{\lambda,\uparrow,\bm{Q+q-k}}^{\dagger}a_{\nu,\downarrow,\bm{k}}^{\dagger}a_{\mu,\downarrow,\bm{q}}|\rm{FS}\rangle_{N}.
\end{align}
where $\phi_{\bm k}^{\mu\nu},\psi_{\bm{Q}}^{\mu},\psi_{\bm{k,q}}^{\mu\nu\lambda}$ are the corresponding wave functions, and $\bm{Q}$ is the center-of-mass momentum. For simplicity, we have dropped higher-order terms in the particle-hole expansions for both states. From the equations $(H'_{0}+H'_{\rm int})|\alpha\rangle_{\bm Q}=(E_{\alpha}+E_{\rm th})|\alpha\rangle_{\bm Q}$ ($\alpha=P,M$), we can derive the closed equations for the molecule and the polaron states. The closed equation for the molecular state takes the form $\det(G')=0$, where the definitions of $G'$ and $F'_{mn}$ are similar to those in Eqs.~(\ref{CEM2}) and (\ref{Fmn}), except that the summation over $\bm{k}$ is constrained by $E_{\bm{k},\downarrow}^{\nu}>E_{F}$ and that $E_{M2}$ is replaced by $\tilde{E}_{M}$. Here, we have $\tilde{E}_{M}=E_{M}+E_{F}+\epsilon_{\uparrow}^{0}$. The closed equation for the polaron state has the form
\begin{align}
\det(KT)+{\rm Tr}(K{\sigma_z}{\sigma_x}T{\sigma_x}{\sigma_z})=1,
\end{align}
where $\sigma_x$ and $\sigma_z$ are the Pauli matrices and the matrix elements of $K$ and $T$ are
\begin{align}
K_{mn}&=\sum_{\mu}\frac{\beta_{\uparrow,\bm{Q}}^{m,\mu}\beta_{\uparrow,\bm{Q}}^{n,\mu}}{(\tilde{E}_{P}-E_{\bm{Q},\uparrow}^{\mu})},\nonumber\\
T_{mn}&=\sum_{\mu,E_{\bm{q},\downarrow}^{\mu}<E_{F}}\frac{\beta_{\downarrow,\bm{q}}^{m,\mu}\beta_{\downarrow,\bm{q}}^{n,\mu}}{\det G'(\tilde{E}_{P}+E_{\bm{q},\downarrow}^{\mu},\bm{Q}+\bm{q})}\nonumber\\
&\times G'_{mn}(\tilde{E}_{P}+E_{\bm{q},\downarrow}^{\mu},\bm{Q}+\bm{q}),
\end{align}
with $\beta_{\sigma,\bm{k}}^{1,\pm}=\sin\theta_{\sigma,\bm{k}}^{\pm}$, $\beta_{\sigma,\bm{k}}^{2,\pm}=\cos\theta_{\sigma,\bm{k}}^{\pm}$, and $\tilde{E}_{P}=E_{P}+\epsilon_{\uparrow}^{0}$.

The energies of the polaron and the molecule state can be obtained by solving the closed equations above and looking for the $\bm Q$ sector with the lowest energy. Here the threshold energies are $E_{\rm th}=\epsilon^0_{\uparrow}+\sum_{E_{\bm{k},\downarrow}^{-}<E_{F}}E_{\bm{k},\downarrow}^{-}$, where $E_F$ is the Fermi energy. For simplicity, we only consider the case where the Fermi sea $|{\rm FS }\rangle$ is entirely in the lower helicity branch. As shown in Fig.~\ref{fig:fig3}, the polaron-molecule transition is now a function of both $B$ and $\Omega$. In the zero-magnetic-field limit, a polaron-molecule transition can be tranversed by tuning the dressing parameter $\Omega$ alone.

\begin{figure}[tbp]
\center{\includegraphics[width=1\linewidth]{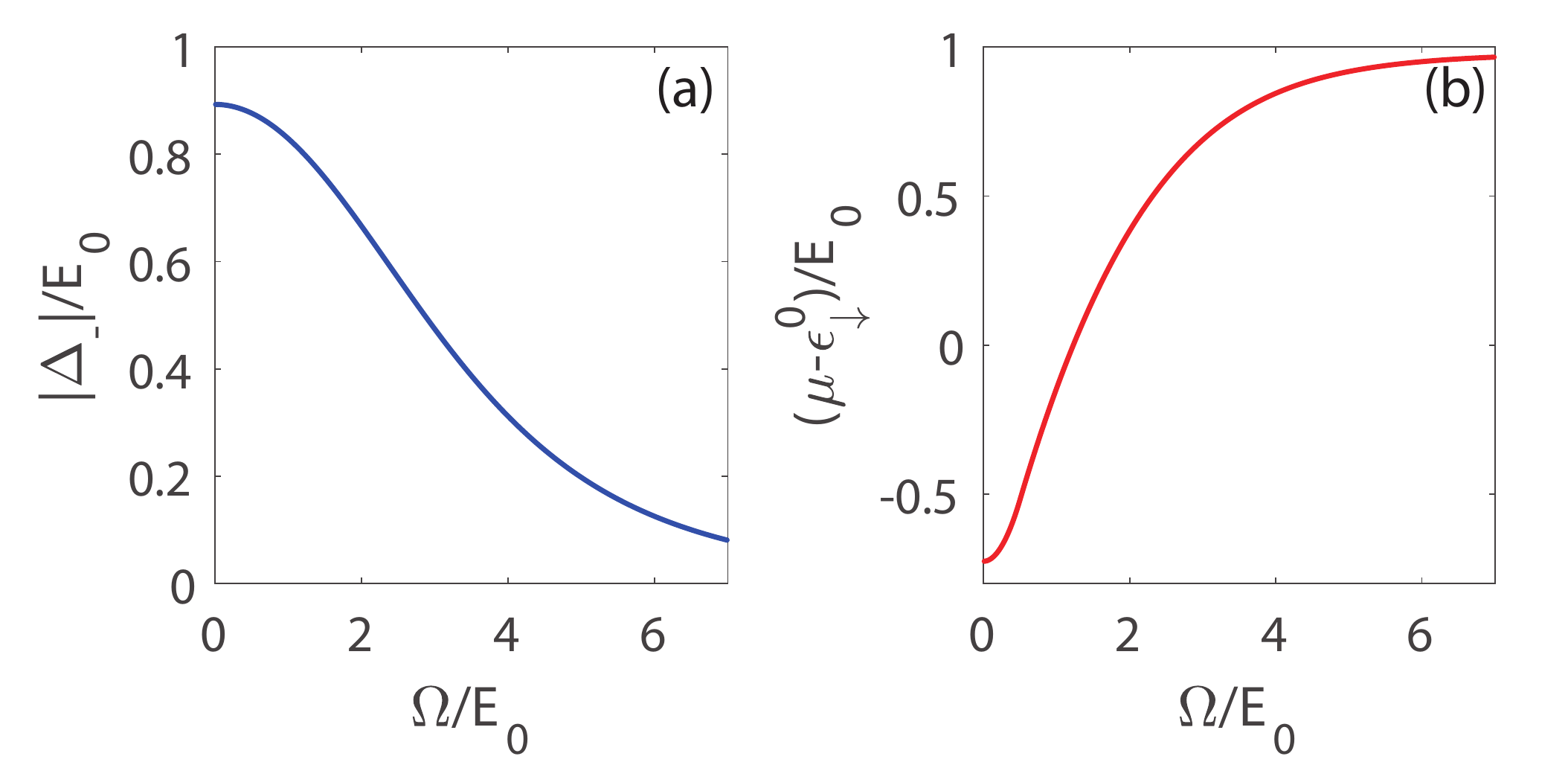}}
\caption{(a) Pairing order parameter $\Delta_-$ as a function of $\Omega$ at $B=0$. Note $\Delta_+=0$ at $B=0$. (b) The chemical potential relative to the single-particle dispersion minimum $\epsilon^0_{\downarrow}$ at $B=0$. We have fixed the total particle density $n=k_{0}^{3}/3\pi^{2}$. The atomic parameters of $^{173}$Yb are shown in the caption of Fig.~\ref{fig:fig2}.}
\label{fig:fig4}
\end{figure}

\emph{BCS-BEC crossover}.--
To demonstrate the impact of dressed FR in the fully many-body environment, we now study the BCS-BEC crossover near an OFR under the Rabi scheme. For simplicity, we will focus on the case of zero magnetic field, and show that by adjusting the dressing parameter, the system can be tuned through the crossover region and into the BCS regime. In the case of non-zero magnetic fields, the asymmetry in the single-particle dispersion induced by the differential Zeeman shifts should give rise to interesting Fulde-Ferrell pairing states~\cite{FF1,FF2}.

Following the BCS-type mean field approach, we write the mean-field interaction Hamiltonian as
\begin{align}\label{intMF}
H_{\rm int}^{\rm MF}&=\left[\Delta_{+}A_{+}^{\dagger}(0)+{\rm H.c.}\right]+\left[\Delta_{-}A_{-}^{\dagger}(0)+{\rm H.c.}\right]\nonumber\\
&-\frac{2}{g_{-}}\Delta_{-}^{2}-\frac{2}{g_{+}}\Delta_{+}^{2},
\end{align}
where the order parameters are defined as $\Delta_{-}=({g_{-}}/{2})\langle A_{-}(0)\rangle$ and $\Delta_{+}=({g_{+}}/{2})\langle A_{+}(0)\rangle$. We have assumed zero center-of-mass momentum for the pairing mean fields, which is consistent with results for the two-body and the molecule states at $B=0$. The effective Hamiltonian  $H_{\rm eff}=H'_{0}+{H}_{\rm int}^{\rm MF}-\mu{N}$ is then
\begin{widetext}
\begin{align}\label{HMF}
H_{\rm eff}=\sum_{\bm{k}}\Psi^{\dagger}(\bm{k})\mathcal{M}(\bm{k})\Psi(\bm{k})-\frac{2}{g_{-}}\Delta_{-}^{2}-\frac{2}{g_{+}}\Delta_{+}^{2}+\sum_{\bm{k}}2(\epsilon_{\bm{k}}-\mu),
\end{align}
where  $\mu$ is the chemical potential, $N$ is the total particle number in the relevant clock states, and
\begin{align}\label{Mk}
&\mathcal{M}(\bm{k})\nonumber\\
&=\left(\begin{array}{cccc}
\epsilon_{\bm{k}}+\frac{k_{0}k_{x}}{2m}-\mu & \Omega & 0 & \Delta_{+}+\Delta_{-}\\
\Omega & \epsilon_{\bm{k}}-\frac{k_{0}k_{x}}{2m}-\mu &  \Delta_{+}-\Delta_{-} & 0\\
0 &  \Delta_{+}-\Delta_{-} & -(\epsilon_{\bm{k}}-\frac{k_{0}k_{x}}{2m}-\mu) & -\Omega\\
 \Delta_{+}+\Delta_{-} & 0 & -\Omega & -(\epsilon_{\bm{k}}+\frac{k_{0}k_{x}}{2m}-\mu)
\end{array}\right),
\end{align}
and $\Psi^{\dagger}(\bm{k})$ is defined as $(\begin{array}{cccc}
a_{e,{\downarrow},\bm{k}}^{\dagger} & a_{g,{\downarrow},\bm{k}}^{\dagger} & a_{e,\uparrow,\bm{-k}} & a_{g,{\uparrow},-\bm{k}}\end{array})$.
\end{widetext}

We then diagonalize ${\mathcal{M}}(\bm{k})$ with the Bogoliubov transformation $\mathcal{M}(\bm{k})X_{\alpha}=E_{\alpha}(\bm{k})X_{\alpha}$ ($\alpha=1,2,3,4$), and obtain the quasi-particle energy $E_\alpha(\bm{k})$ together with the vectors of Bogoliubov coefficients $X_{\alpha}$. This leads to the zero-temperature thermodynamic potential ${\mathcal K}=\langle{H_{\rm eff}}\rangle_{\rm BCS}$ (here the expectation value is taken with respect to the BCS ground state)
\begin{equation}\label{Kmfdia}
  {\mathcal K}=\sum_{\bm k \alpha}\Theta(-E_{\alpha}(\bm k))E_{\alpha}(\bm{k})+\sum_{\bm{k}}{2(\epsilon_{\bm k}-\mu)}-\frac{2}{g_{-}}\Delta_{-}^{2}-\frac{2}{g_{+}}\Delta_{+}^{2},
\end{equation}
where $\Theta(x)$ is the Heaviside step function. The gap and the number equations can be obtained, respectively, from the conditions $\partial {\mathcal K}/\partial \Delta_{\pm}=0$, and $\partial {\mathcal K}/\partial\mu=-N$, from which it is straightforward to solve $\Delta_{\pm}$ and $\mu$.

In Fig.~\ref{fig:fig4}(a), we see that as the dressing parameter $\Omega$ increases, the pairing mean field $\Delta_-$ decreases monotonically, which suggests that the system is approaching the BCS regime. This picture is confirmed in Fig.~\ref{fig:fig4}(b), where the chemical potential relative to
single-particle dispersion minimum $\epsilon^{0}_{\downarrow}$ is shown. With increasing $\Omega$, the relative chemical potential changes its sign from negative to positive, and eventually approaches $E_0$. As $E_0$, by definition, is the Fermi energy of a non-interacting two-component Fermi sea (see Fig.~\ref{fig:fig4} caption), the behavior of the chemical potential is a clear signature that the system changes from bosonic to fermionic, and reaches the deep BCS regime in the large-$\Omega$ limit.

\emph{Final remarks}.--
We have shown that by coupling atomic modes in the two relevant scattering channels of a Feshbach resonance, the resonance position can be made sensitively dependent on the coupling parameters. This provides further tunability to prepare atoms in the strongly interacting regime. In light of the recent experimental realization of spin-orbit couplings in alkaline-earth atoms, our prediction can be readily observed experimentally, and offers exciting possibilities for the quantum simulation using alkaline-earth-like atoms.

\emph{Acknowledgments}.--
This work is supported by the National Key R\&D Program (Grant No. 2016YFA0301700), the NKBRP (2013CB922000), the National Natural Science Foundation of China (Grant Nos. 60921091, 11274009, 11374283, 11434011, 11522436, 11522545, 11774425), and the Research Funds of Renmin University of China (10XNL016, 16XNLQ03). W.Y. acknowledges support from the ``Strategic Priority Research Program(B)'' of the Chinese Academy of Sciences, Grant No. XDB01030200.

\end{document}